# Efficient Task Mapping for Manycore Systems

Xiqian Wang, Jiajin Xi, Yinghao Wang, Paul Bogdan, Shahin Nazarian
*Department of Electrical and Computer Engineering, University of Southern California, Los Angeles, CA, USA*

**Abstract**— System-on-chip (SoC) has migrated from single core to manycore architectures to cope with the increasing complexity of real-life applications. Application task mapping has a significant impact on the efficiency of manycore system (MCS) computation and communication. We present WAANSO, a scalable framework that incorporates a Wavelet Clustering based approach to cluster application tasks. We also introduce Ant Swarm Optimization (ASO) based on iterative execution of Ant Colony Optimization (ACO) and Particle Swarm Optimization (PSO) for task clustering and mapping to the MCS processing elements. We have shown that WAANSO can significantly increase the MCS energy and performance efficiencies. Based on our experiments on a 64-core system, WAANSO improves energy efficiency by 19%, compared to baseline approaches, namely DPSO, ACO and branch and bound (B&B). Additionally, the performance improves by 65.86% compared to Density-Based Spatial Clustering of Applications with Noise (DBSCAN) baseline.

## I. INTRODUCTION

With the development of modern semiconductor technologies, a large number of sources including intellectual property (IP) and embedded memory blocks can be integrated by the designers on a very tiny scale. However, the growing computational resources require tremendous communication volume [1]. Also, the maximum operating frequency of a single-core processor has hit a ceiling due to power dissipation limits and other scaling rules such as short channel effects. This demands SoC designers to pursue parallel computing in MCS designs. As a promising interconnect infrastructure, NoC has a significant impact on mitigating the challenges related to on-chip communication and heterogeneity of cores. However, the full potential of MCS can be realized only if the applications are suitably parallelized considering task dependencies as well as the timing and power constraints [2]–[6]. Applications can be either converted from an existing sequential form or written to be executed in parallel from scratch. Once the code is compiled, it must be mapped efficiently onto the underlying hardware. Many decisions need to be made while mapping and scheduling a program onto an MCS. These include determining how much of the potential parallelism should be exploited, the number of processors to use, how parallelism should be scheduled, etc. The best mapping choice depends on the relative costs of computation, energy consumption, and other hardware and varies from one MCS to another. This mapping can be performed manually by the programmer or automatically by the compiler or run-time system.

There have been numerous works related to task scheduling and mapping, including those for NoCs. E.g., an ACO-based approach minimizes the energy and the thermal power of NoCs [7]. ACO, however suffers from its low convergence because of its unknown initial pheromone on the map. The algorithm of [8] utilizes B&B to achieve a balance between reliability and energy, but it is difficult to set lower bound and upper bound accurately. DPSO was proposed to address IP placement with some augmentations [9]. Unfortunately, DPSO can get trapped in a local optimum and therefore may miss the global optimum. More recently there has been a growing surge in using Machine Learning (ML) for task scheduling and parallelism [10]–[12].

We propose WAANSO, a framework that performs task mapping for multiple applications onto an MCS while optimizing for energy, performance and the required number of cores. Wavelet Clustering finds the minimum number of cores needed to run the given multiple applications with the minimum performance requirement. Our framework also introduces Ant Swarm Optimization (ASO) which is based on iterative execution of the PSO and ACO, in order to map the clusters of tasks provided by the Wavelet Clustering onto the MCS such that the overall performance and energy are optimized.

## II. WAANSO FRAMEWORK

Ideally, the best mapping onto an MCS should provide the minimum total energy cost and execution time. However a shorter execution time may mean higher energy cost and vice versa. Also, shortest execution time for one application may mean, longer execution times for some other applications. Our goal is therefore to find the best mapping based on three constraints: energy consumption, performance of one application, and performance of all applications.

WAANSO acts on the three constraints using two models. One deals with the performance of applications independently and collectively. Another model provides the balance of energy and performance within one application. The basic flow of WAANSO is shown in Fig. 1. We first use Wavelet Clustering to find the minimum number of cores to run the given applications while minizing performance loss. Next ASO iteratively finds the task cluster mapping onto MCS cores.

### A. Graph Modeling

- Each application with n tasks is represented as $CG < C, W >$, which is referred as the communication weighted graph [13]. $C$ is the set of n vertices, each representing a task. $W$ is the set of edge weights, each modeling the communication volume between the two corresponding nodes for $M$ applications. our framework deals with $M$ $CG < C, W >$ graphs, i.e., one graph per application.

- The MCS is modeled by an MCS topology graph $TG < R, L >$, where $R$ is the set of routers, and $L$ is the set of links in the graph. $l_{i,j}$ is the link between router $r_i$ and $r_j$. $b_{i,j}$ is the bandwidth of the link. The problem is then transformed to mapping of $M$ $CG < C, W >$ onto $TG < R, L >$ aiming for optimization of a unified energy and performance cost.

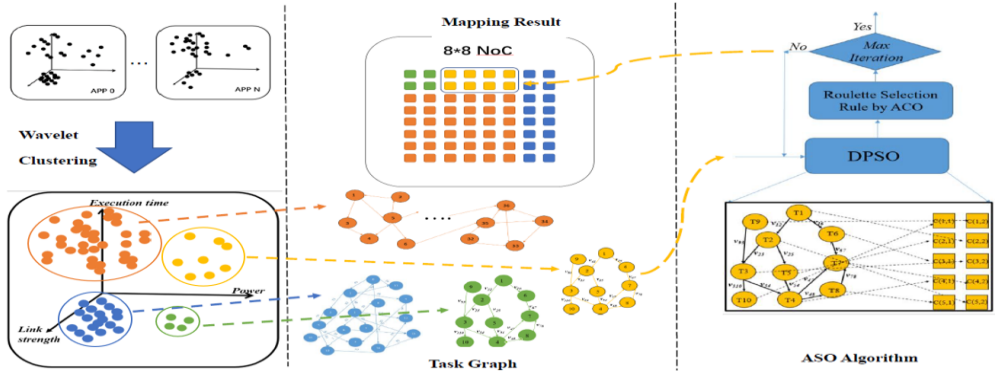

Fig. 1. Framework of WAANSO

## B. Energy Model

Bit energy $E_{bit}$, is used to estimate the dynamic energy consumption for each bit [1]. The following equation computes the dynamic energy consumed by all bits, $E_{bit}^{\tau_i,\tau_j}(m)$, traversing the NoC from tile $\tau_i$ to $\tau_j$ passing through n routers in MCS:

$$E_{DyNoc}^{\tau_i,\tau_j} = \sum_{m=1}^{n} E_{Bit}^{\tau_i,\tau_j}(m) \qquad (1)$$

## C. Performance Model

The cost function of performance can be described as:

$$cost_{perf}^{a} = best\{f_{cost}\} \qquad (2)$$

$f_{cost}$ is a function to calculate the best result of trade-off between the number of executing cores and execution time. $Cost_{perf}$ is the best trade-off result of performance and energy defined by adding up the minimum $f_{cost}$ for every application [14].

## D. Unified Performance and Energy Cost Model

Using a hyper parameter $\alpha$, the objective cost function that measures the tradeoff between energy and performance is:

$$Cost_{total} = \alpha \frac{Cost_{perf}}{\max Cost_{perf}} + (1-\alpha)\frac{Cost_{ener}}{\max Cost_{ener}} \qquad (3)$$

## III. WAVELET CLUSTERING

### A. Wavelet Clustering

Wavelet Clustering is efficient in managing cases with sparse data in multidimensional feature spaces, high orders and intense noise of input data [15]. The data $X$ for each application is related to the number of idle cycles of the core, and the data throughput per cycle [16]. $X=x_1, x_2, ...x_P$ denotes the data for an MCS with P cores. Applying Mallat algorithm by using Discrete Wavelet Transform (DWT) $j$ times, the features of $x_i$ are defined by the general feature, $c_k^j$, and the detailed feature, $d_k^j$, with the wavelet coefficient $k$ during this period [17] as follows:

$$c_k^j = \sum_{x=0}^{N-1} lo(x) c_{x+2k}^{j-1}, j \in J \qquad (4)$$

$$d_k^j = \sum_{x=0}^{N-1} hi(x) c_{x+2k}^{j-1}, j \in J \qquad (5)$$

$lo(x)$ and $hi(x)$ are low and high pass filters, respectively, and N is the length of filter operator.

Next, the cycle-based behavior information of each core is obtained. To cluster similar cores together, we use the following DWT steps based on the steps explained in [18], namely Quantization, DWT, Significant Grid Identification, and Cluster Identification. The outcome of these steps would be the minimum number of cores to run the application. However, one weakness of Wavelet Clustering is that the connections between clusters may cause modifiable areal unit problem (MAUP) [18]. This means the required number of connections is so high, that few clusters will be found. We therefore recognize the strength of these connections between clusters as part of our WAANSO.

### B. Strength of Link

*a) Related Wavelet Entropy (RWE).* Using DWT, the WE for input $x \in X$ is calculated using discrete wavelet $D_j(k)$ [19]:

$$WE(k,j)=[-\sum_j P_{D(j,k)}\ln P_{D(j,k)}, -\sum_k P_{D(j,k)}\ln P_{D(j,k)}] \qquad (6)$$

$P_{D(j,k)}$ is the disorder or complexity that input data presents at the $j^{th}$ transform using a wavelet with scale $k$ [19], [20]. $P_{D(j,k)}$ can be replaced by $c_k^j$ or $d_k^j$ [20], therefore:

$$WE(k,j) = [-\sum_j c_k^j \ln c_k^j, -\sum_k d_k^j \ln d_k^j] \qquad (7)$$

RWE for $p, q \in X$ is [18]:

$$RWE(p|q) = \left[\sum_j p_{c_k^j}\ln\left(\frac{p_{c_k^j}}{q_{c_k^j}}\right), \sum_k p_{d_k^j}\ln\left(\frac{p_{d_k^j}}{q_{d_k^j}}\right)\right] \qquad (8)$$

*b) Strength of Connection.* We define the link strength $s_{(i,j)}$ of clusters $x_i$ and $x_j$ based on the following :

$$s_{(i,j)} = \sum_a^i \sum_b^j (RWE(a|b)), a,b \in c_{(i,j)} \qquad (9)$$

$s_{exp}$ is self-adapted to find the most suitable value for the current clustering results. The self-adapting process is similar to the update method of self-adapting filter [21] and neural networks. $s_{exp}$ is calculated per cycle and updated based on (9) and (10) below [17], [22]:

$$Set \quad s_{exp}^{initial} = \sum_a^{x_i} \sum_b^{x_j} \left(\frac{1}{a_{c_k^j} b_{c_k^j}}(var(a)var(b))\left(a_{d_k^j} b_{d_k^j}\right)\right)$$

$$s'_{exp} = \omega \cdot s_{exp} + \varepsilon s_{(xi,xj)} \qquad (10)$$

$s'_{exp}$ represents $s_{exp}$ in the next cycle. (11) is wavelet variance [23], [24]:

$$var(x) = \sum_k \left(\sum_j x_{d_k^j}\right)^2 \tag{11}$$

$\varepsilon$ is the recursive least squares error [4] from initial to current cycle [21], [25]:

$$\varepsilon = \sum_j \in (current - 1)((s_{exp}^j - s'^j_{exp})^2) \tag{12}$$

and the updated weight vector of $s_{exp}$, $\omega'$ in each cycle is [22]:

$$\omega' = \omega + \mu\gamma \left[\sum_{j \in current} \sum_k \left(a_{c_k^j}\right)^2 \left(b_{c_k^j}\right)^2\right]^{\frac{1}{2}} \tag{13}$$

$\gamma$ is the learning rate randomly selected from range $1 - 2\gamma(MEANa_j^k + MEANa_j^k)^2 \in [0, 1]$. $\mu$ is the forgotten factor which is a positive number less than 1 and updated as [21], [23]:

$$\mu' = \frac{1}{\sum_{j \in current} 2\left(MAX\{a_{d_k^j}\} + MAX\{b_{d_k^j}\}\right)^2} \tag{14}$$

We can use $s_{exp}$ as the expected link strength when $\varepsilon$ is minimized. If $s_{(i,j)} < s_{exp}$, the connection is considered weak and hence will be ignored. Hence we can get a more accurate result of minimum number of cores since fewer cores are mis-located.

*C. Performance vs. Number of Clusters Trade-off*

Performance is a nonlinear function of the number of cores in MCS and saturates as the number of cores is at a certain threshold [26]. Furthermore, increasing the number of cores cannot help the performance or energy consumption when it exceeds the number of task executing paths. The number of task clusters is related to the size of NoC grids, i.e., a large grid will give few clusters, thus the data with different features may be in the same cluster and vice versa. To find the minimized grid size that will cluster data more optimally, we first make the grid as large as possible and calculate $cost_{perf}$. We then increase the number of grids to a point where $cost_{perf}$ is saturated. $f_{cost}$, in Eqn. (3), is the reciprocal of sum for n clusters' self-entropy:

$$f_{cost} = \frac{1}{\sum_{i=1}^{n} \frac{c_i}{m_d} H_i} \tag{15}$$

where $c_i$ is the number of data in cluster $i$ and $m_d$ is the number of total data. $H_i$ the cluster $i$ entropy or diversity of feature of data in the cluster, is entropy sum of all data, $p_j$, in the cluster:

$$H_i = -\sum_{j \in c_i} p_j \ln p_j \tag{16}$$

$$p_j = \frac{\sum_k |d_k^j|^2}{\sum_{j \in J} \sum_k |d_k^j|^2} \tag{17}$$

IV. ANT SWARM OPTIMIZATION (ASO)

PSO is a population-based intelligence optimization algorithm [27] where each particle has a location vector $Xk = <x_{k,1}, x_{k,2} \cdots x_{k,n}>$ in $k^{th}$ iteration. The velocity vector of a particle is $V_i^k = (v_1^k, v_2^k \cdots v_n^k)$, which corresponds to the swapped sequence of two different mapping results. Particle $i$ in iteration $k$ updates its mapping result base on the following:

$$V_i^{k+1} = \omega V_i^k + c_1 r_1 (pBest_i^k - X_i^k) + c_2 r_2 (gBest_i^k - X_i^k) \tag{18}$$

$$X_i^{k+1} = X_i^k + V_i^{k+1} \tag{19}$$

$r_1$ and $r_2$ are two random numbers distributed between 0 and 1, $c_1$ and $c_2$, are positive acceleration constants, $pBest_i^k$ is the best mapping result of particle $i$ in $k^{th}$ iteration, and $gBest_i^k$ is the global best mapping result found from $k$ iterations. The total cost function $Cost_{total}$ is used as the fitness value to evaluate the mapping result of one particle. The properties of the particle are adjusted by all particles' experiences.

The diversity, namely the ability of preventing the solution falling into local optimum, can be increased by changing the global and local optimum fitness value in each iteration, and applying the random number and the acceleration parameter according to ACO. In ACO, the next node is determined by an ant based on the pheromone on the map. The idea of improving swarm optimization by using the roulette selection rule in ACO was presented in [28]. The particle's velocity is updated according to the corresponding global and local optimum values, $pBest_i^k$ and $gBest_i^k$. We therefore store these values after each iteration of PSO. The next global optimum fitness value, $F_{gb}(t+1)$, is calculated based on the following transition rule:

$$F_{gb}(t+1) = \begin{cases} \min_{j \in [1,t]}(F_{gb,j}) & , q < q_0 \\ \dfrac{F_{gb,j}}{\sum_{j=1}^{t} F_{gb,j}} & , q \geq q_0 \end{cases} \tag{20}$$

$F_{gb}$ is the global fitness value, $t$ is an iteration number, $q$ is a random number, and $q_0$ is the threshold value based on experience. The particle moves to the current optimum and non-optimum fitness value, which increases the diversity of a particle. After improving PSO, diversity is increased and the probability of a particle falling into the local optimum is reduced. In a typical setup for PSO, position update at iteration $t+1$ is based on the location and velocity values at iteration $t$. To increase the converging speed of particles, the location vector is updated based on the global optimum position value. A random velocity, $rand(v)$ has been introduced to increase the diversity:

$$x_i(t+1) = \begin{cases} x_i(t) + v_i(t+1) & , q' \geq q_1 \\ p_{gb}(t) + rand(v) & , q' < q_1 \end{cases} \tag{21}$$

$q_1$ is the threshold value based on experience, $q'$ is a random number to compare with the threshold to determine position update, and $p_{gb}(t)$ is global best of particle p in the $t^{th}$ iteration.

V. EXPERIMENTAL RESULTS

Three mapping techniques, namely B&B, ACO, and PSO were used as baselines to evaluate ASO. Also DBSCAN [16] was used to evaluate Wavelet Clustering of WAANSO. All experiments were simulated on a Linux system with a 2.6 GHz Intel Core i7-6700HQ processor with 16GB memory. Wavelet Clustering was implemented in MATLAB. We use the B&B of [1] and DBSCAN of scikit-learn [29]. ACO, DPSO, and ASO were implemented in C++. All algorithms were packaged in

Python. Although our WAANSO is applicable to heterogeneous MCS and various typologies, we used a mesh-based 8x8 NoC of (64) homogeneous cores for simplicity. We also utilized a cycle-based NoC simulator called NOXIM [30], whose energy data was modified with data from [8] to fit B&B. We used data and model of eight applications from real applications in [26].

*A. Performance Evaluation*

We first run each application separately using all the algorithms listed in Table 1. ASO outperforms B&B, ACO and PSO. Also Wavelet Clustering increases performance, e.g., by 65% in case of WAANSO, whereas DBSCAN based baselines are not as effective.

*B. Energy Dissipation Evaluation*

ASO consistently generates the best mapping result with the smallest energy consumption (Fig. 2), e.g., about 23.5% lower than that of B&B. Also results of Fig.3 confirm that WAANSO produces the best mapping in terms of energy consumption.

TABLE 1. PERFORMANCE EVALUATIONS

| Algorithm | Executing Time(No. clock cycles) |
|---|---|
| B&B only | 1476 |
| ACO only | 1481 |
| PSO only | 1475 |
| ASO only | 1439 |
| WC + B&B | 579 |
| WC + ACO | 592 |
| WC + PSO | 583 |
| WAANSO | 567 |
| DBSCAN + B&B | 1101 |
| DBSCAN + ACO | 1697 |
| DBSCAN + PSO | 1683 |
| DBSCAN + ASO | 1661 |

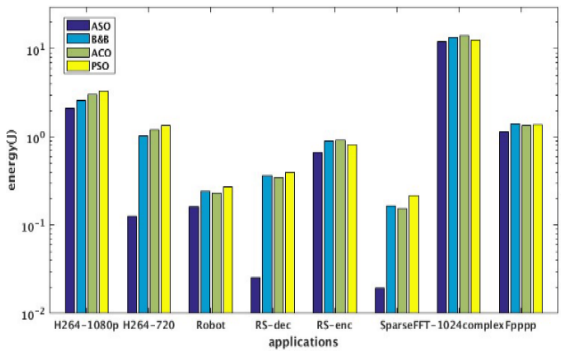

Fig. 2. Total energy with mapping algorithms only.

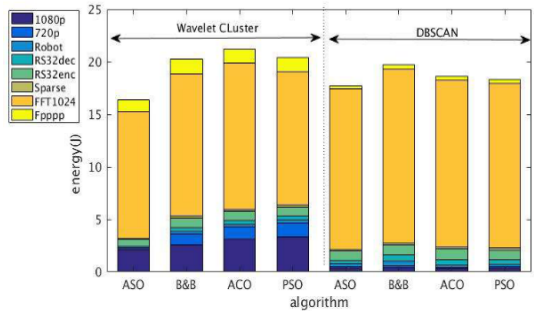

Fig. 3. Total energy with WC and DBSCAN.

## VI. CONCLUSION

We presented WAANSO, an iterative MCS task mapping framework based on wavelet clustering, ACO and PSO that significantly improves performance and energy efficiencies compared to B&B, PSO, ACO and DBSCAN baselines.